\documentclass[prl,aps,twocolumn,showpacs]{revtex4}
 \newcommand{\mytitle}[1]{
 \twocolumn[\hsize\textwidth\columnwidth\hsize
 \csname@twocolumnfalse\endcsname #1 \vspace{1mm}]}
 \newcommand{\beq}{\begin{equation}}
 \newcommand{\eeq}{\end{equation}}
 \newcommand{\bea}{\begin{eqnarray}}
 \newcommand{\eea}{\end{eqnarray}}

\usepackage{graphics}
\usepackage{graphicx}
\usepackage{amssymb}
\usepackage{color}
\usepackage{bm}

\begin{document}

\title{Gate-tuned two-channel Kondo screening by graphene leads:\\
 Universal scaling of the nonlinear conductance}
\author{Tsung-Han Lee$^{1}$, Kenneth Yi-Jieh Zhang$^{1}$, 
Chung-Hou Chung$^{1,2}$, Stefan Kirchner$^{3,4}$}
\affiliation{
$^{1}$Electrophysics Department, National Chiao-Tung University,
HsinChu, Taiwan, 300, R.O.C. \\
$^{2}$National Center for Theoretical Sciences, HsinChu, Taiwan, 300, R.O.C. \\
$^{3}$Max-Planck-Institut f\"ur Physik komplexer Systeme, 01187 Dresden, Germany\\
$^{4}$Max-Planck-Institut f\"ur chemische Physik fester Stoffe, 01187 Dresden, Germany
}

\date{\today}

\begin{abstract} 
Based on the non-crossing approximation,
we calculate both the linear and nonlinear conductance within the 
two-lead two-channel single-impurity Anderson model where the 
conduction electron 
 density of states vanishes in a power-law fashion
$ \propto |\omega-\mu_F|^r$ with $r=1$ near the Fermi energy, appropriate for an hexagonal system. 
For given gate voltage, we address the universal crossover
from a two-channel Kondo  phase, argued to occur in doped graphene, 
to an unscreened local moment  phase. 
We extract universal scaling 
functions in conductance governing charge transfer through 
the two-channel pseudogap Kondo impurity and discuss our results in the context of a  
recent scanning tunneling spectroscopy experiment on Co-doped graphene.
\end{abstract}

\maketitle
{\it Introduction.}
\begin{figure}[t]
\begin{center}
\includegraphics[angle=-90,width=0.8 \linewidth]{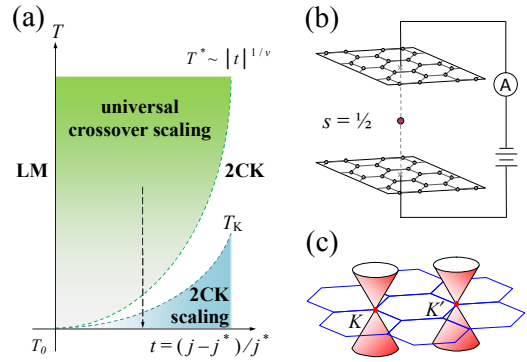}
\end{center}
\par
\vspace{-0.5cm}
\caption{(a) Schematic phase diagram for the 2CK-LM crossover. 
The parameter 
$j$ refers to either $\Gamma$ or $\mu$ (in units of $D=1$), 
and $j^{\ast}$ refers to the crossover scale for a fixed temperature 
$T_0\sim 5\times 10^{-7} D$. 
$T_K$ and $T^{\ast}$ represent  crossover energy scales 
associated with the universal scaling for 2CK 
(blue shaded) and the high-temperature 2CK-LM crossover scaling 
(green shaded) regime, respectively. 
(b) Schematic setup of the model in Eq. (1): a $S=1/2$ impurity 
(red dot) couples symmetrically to the two sub-lattices of two 
graphene leads that are kept at different chemical potentials 
of the setup. (c) Dirac points 
(labeled by $K$ and $K^{\prime}$) with its Dirac spectrum on graphene's honeycomb lattice 
in momentum space.  
}
%\vspace{-2cm}
\label{phase}
\end{figure} 
The two-channel Kondo~\cite{kondo} (2CK) problem~\cite{2CK} 
is a fascinating example of an exotic quantum many-body phenomenon resulting in a metallic ground state
with non-Fermi-liquid behavior. It involves a single quantum impurity 
spin with $s=1/2$ that couples antiferromagnetically to two identical
conduction electron reservoirs. As a result, Kondo processes involving both reservoirs lead to overscreening 
of the local moment.
Theoretically, the 2CK physics has been studied extensively
via Bethe ansatz~\cite{Bethe}, conformal field theory~\cite{CFT}, 
bosonization~\cite{bosonization} and the numerical renormalization group~\cite{2CK-NRG}. 
Experimentally, however, up to date, only very few examples 
of clear 2CK physics have been experimentally
realized, {\it e.g.} in semiconductor quantum dots~\cite{Goldhaber}, in 
magnetically doped nanowires, and in metallic glasses~\cite{Cox_a,Cichorek}. 

Recently, the Kondo effect of magnetic adatoms in graphene has 
attracted much attention, 
theoretically~\cite{sengupta,Vojta2} as well as 
experimentally~\cite{Mattos} due to the possible 
realization of a 2CK ground state.
One interesting aspect of Kondo physics in graphene is 
due to the Dirac (linear) spectrum that gives raise to
a pseudogap local density of states (DOS), 
$\rho_c(\omega) \propto |\omega|^r$ with $r=1$, at the impurity site, 
making graphene one of the few experimental realizations of 
the pseudogap Kondo model~\cite{Fradkin,Gonzalez-Buxton.98}, 
the simplest model to study critical Kondo 
destruction~\cite{Ingersent,Glossop_a}.
In the pseudogap Kondo (or more generally Anderson) model, 
a quantum phase transition 
is expected between the Kondo screened 
and the unscreened local moment (LM) ground states for $0<r<1$~\cite{Gonzalez-Buxton.98, Vojta4}. 
For $r=1$, the case of graphene, Kondo screening does not occur~\cite{Gonzalez-Buxton.98,Vojta2,Vojta3}, resulting in a LM ground state. Yet, Kondo screening 
can be induced by changing the Fermi energy, {\it e.g. } by applying a gate voltage ($\mu\neq 0$).

That two independent screening channels can exists in graphene 
is related to the existence of 
two inequivalent Dirac points 
($K$ and $K^{\prime}$) in its band structure. 
It therefore was argued that the effective low-energy model for 
magnetic impurities in 
graphene depends on the location of the adatom~\cite{DAnna}: if the 
impurity is located at the center of the cell, the inter-valley 
scattering does not couple the two screening channels and an effective 
2CK ensues~\cite{sengupta}. This has been explicitly demonstrated within a 
tight-binding description where the hybridization between electronic states 
in graphene and impurity states preserves the A-B sub-lattice 
symmetry~\cite{GZhu}. 
The situation is different if the 
adatom is located on a graphene site and
the sub-lattice symmetry is absent in the impurity-graphene hybridization. 
The effective low-energy model is in this case
the more conventional one-channel pseudogap Kondo 
model~\cite{DAnna,GZhu}. The difference between the various adatom positions can be probed by scanning tunneling microscopy (STM)~\cite{Wehling,note}. This has been achieved recently 
with $Co$-atoms as magnetic adatoms where
signatures of 2CK physics are seen when the adatom is located at the zone 
center of the honeycomb cell  such that the inter-valley scatterings between 
$K$ and $K^{\prime}$ Dirac electrons are strongly suppressed~\cite{Mattos}.

\begin{figure}[t]
\begin{center}
\includegraphics[width=0.8\linewidth]{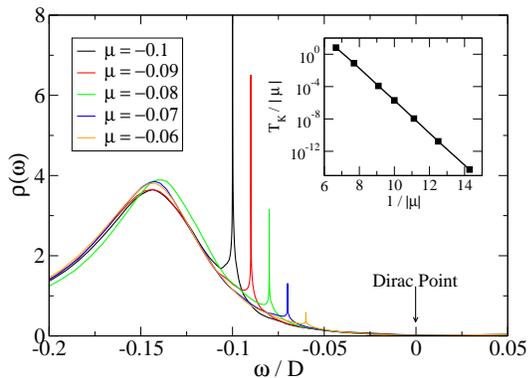}
\end{center}
\par
\vspace{-0.5cm}
\caption{The impurity spectral function $\rho(\omega)$ vs.
chemical potential $\mu$ (in units of the bandwidth $D$). 
The Kondo peak is pinned near $\mu$. The Dirac point is assigned $\omega=0$. 
Inset: Kondo temperature $T_K$ (defined in text) as a function of $\mu$. 
The NCA parameter are $T=5\times10^{-7}D$, $\Gamma=0.2D$, $\epsilon_d=-0.2D.$}
\vspace{0.5cm}
\label{DOSfig}
\end{figure}

The 2CK-LM quantum phase transition  
in the two-channel pseudogap Kondo and Anderson model 
with $\mu=0$ and $0<r<1$ in and out of equilibrium has been 
studied recently~\cite{Gonzalez-Buxton.98,Vojta3,Zamani.12,Zamani.13}, but 
the experimentally more relevant situation of
$r=1$  
and $\mu\neq 0$ 
has not yet been properly addressed. 
The possibility of realizing 2CK physics was first pointed out in 
Ref.~\cite{sengupta}. There is, however, a lack of systematic  
investigations beyond the mean-field treatment. 
In this letter, we study the crossover phenomenon from the LM to 2CK regime,
and work out the  universal scaling functions of the conductance. 
A comprehensive analysis of the  non-equilibrium transport, including  
STM lineshapes in the various regimes is presented.
We address this issue 
using the non-crossing approximation 
(NCA)~\cite{Cox_b,Meir,kroha}, which is known to give reliable results for 
multi-channel Kondo systems which are in line with conformal field theory results~\cite{Cox_a}.
It has been shown recently that the NCA is 
able to reproduce the correct qualitative features of the two-channel 
pseudogap model for $0\leq r \leq 1$ in and out of equilibrium in excellent agreement with exact results~\cite{Zamani.12,Zamani.13}.

Here, we study the steady-state transport properties of the 
experimentally most relevant pseudogap case, $r=1$ as a function of doping, temperature and bias voltage.

{\it The Model Hamiltonian.} 
Our starting point  
is the two-lead two-channel single-impurity pseudogap Anderson model. Each lead is characterized by
a  power law density of states (DOS) ($\rho_{c}(\omega)\sim|\omega|\Theta(D-|\omega|)$, 
where $D$ is a high-energy cutoff that 
serves as our unit of energy $D=1$).
A possible  
realization of this model is  sketched in Fig. 1 (b) 
where a spin-$1/2$ impurity is coupled to two  graphene leads with 
couplings to each lead preserving the sub-lattice symmetry of the 
honeycomb lattice~\cite{GZhu,Mattos}.

Within a pseudofermion
representation, the Hamiltonian reads~\cite{Meir,kroha}
\begin{eqnarray}
H &=& \sum_{k, \sigma, \tau, \alpha} (\epsilon_k-\mu_\alpha) c^{\alpha \dagger}_{k\sigma\tau} c^{\alpha}_{k\sigma\tau} + 
\epsilon_{d}\sum_{\sigma} f^{\dagger}_{\sigma} f_{\sigma} \nonumber \\
 &+& \sum_{k, \sigma, \tau, \alpha}\big(U_{\alpha}(f^{\dagger}_{\sigma}b_{\tau}c^{\alpha}_{k,\sigma,\tau})
+\mbox{h.c.}\big),
\label{Hamiltonian}
\end{eqnarray}
where $\mu_\alpha$
%= \pm \mu\equiv \pm \frac{eV}{2}$ being the chemical 
is the chemical 
potential of lead $\alpha$, and $\alpha=L/R$ labels the two conduction electron leads. The indices $\sigma$ and $\tau$ refer to spin and charge (related to K and K$^{'}$) channels characterizing the conduction electrons. The second 
and third term of the RHS of Eq.~(\ref{Hamiltonian}) 
represent the spin-$1/2$ and the hybridization strength $U_{\alpha}$ between the graphene electrons
and the impurity. In the pseudofermion representation the local electron is decomposed
as $d^{\dagger}_{\sigma,\tau}=f^{\dagger}_{\sigma}b_{\tau}$. 
Eq.~(\ref{Hamiltonian}) is a faithful representation of the two-channel single-impurity pseudogap Anderson model, 
provided the
 constraint $Q=\sum_{\sigma}f^{\dagger}_{\sigma}f_
{\sigma}+\sum_{\tau}b^{\dagger}_{\tau}b_{\tau}=1$ is fulfilled at all times.
A finite bias voltage is implemented by shifting the chemical potentials in the leads such that
$\mu_L-\mu_R=eV$ is the applied bias voltage across the 2CK system~\cite{note}.

To study the properties of Eq.~(\ref{Hamiltonian}) we 
employ the NCA. Its ability to correctly capture the  properties of the pseudogap two-channel Anderson model was established recently~\cite{Zamani.12,Zamani.13}.
Within the NCA, the retarded self-energy for pseudofermions, $G^{r}(\omega)=[\omega-\epsilon_{d}-\Sigma^r(\omega)]^{-1}$, and slave-bosons, $D^{r}(\omega)=[\omega-\Pi^{r}(\omega)]^{-1}$, are~\cite{kroha,Meir,Zamani.12,Zamani.13}
\begin{eqnarray}
\Sigma^{r}(\omega)=\frac{2}{\pi}\sum_{\alpha}\int d\epsilon
\Gamma_{\alpha}(\omega-\epsilon-\mu_\alpha)f(\epsilon-\omega-\mu_\alpha)D^{r}(\epsilon),
\label{SFeq1}
\end{eqnarray}
\begin{eqnarray}
\Pi^{r}(\omega)=\frac{2}{\pi}\sum_{\alpha}\int d\epsilon
\Gamma_{\alpha}(\epsilon-\omega-\mu_\alpha) 
f(\epsilon-\omega-\mu_\alpha) G^{r}(\epsilon).
\label{SFeq2}
\end{eqnarray}

The NCA expressions for the lesser self-energy of the pseudofermion, 
$G^{<}(\omega)=\Sigma^{<}(\omega)|G^{r}(\omega)|^{2}$, and slave-boson, $D^{<}(\omega)=\Pi^{<}(\omega)  |D^{r}(\omega)|^{2}$, are

\begin{eqnarray}
\Sigma^{<}(\omega)=\frac{2}{\pi}\sum_{\alpha}\int d\epsilon
\Gamma_{\alpha}(\omega-\epsilon-\mu_\alpha)
f(\omega-\epsilon-\mu_\alpha) D^{<}(\epsilon),
\label{SFeq3}
\end{eqnarray}
\begin{eqnarray}
\Pi^{<}(\omega)=\frac{2}{\pi}\sum_{\alpha}\int d\epsilon
\Gamma_{\alpha}(\epsilon-\omega-\mu_\alpha)
f(\omega-\epsilon+\mu_\alpha) G^{<}(\epsilon).
\label{SFeq4}
\end{eqnarray}
Here, $\Gamma_{\alpha}(\omega)\equiv \Gamma_\alpha \rho_{c,\alpha}(\omega)$ 
with $\Gamma_\alpha= \pi|U_{\alpha}|^{2}$ with
$\rho_{c,\alpha}(\omega) \propto |\omega-\mu_{\alpha}|$
%$\rho_c(\omega)=|\omega|$ is the power law DOS of graphene for symmetry coupling ($U_{L}=U_{R}$), 
and $f(\omega)=[1+e^{\beta \omega}]^{-1}$ is the Fermi function. 

The physical spectral 
function, $\rho(\omega,V)$, is the convolution of pseudofermion and slave-boson Greens function
\begin{eqnarray}
\rho(\omega,V)=\frac{i}{2\pi^{2}Z}\int d\epsilon[ImD^{r}(\epsilon)G^{<}(\omega+\epsilon)-\nonumber\\D^{<}(\epsilon)ImG^{r}(\omega+\epsilon)].
\label{DOSeq}
\end{eqnarray}

The normalization factor $Z=\frac{i}{\pi}\int d\omega[D^{<}(\omega) \\-G^{<}(\omega)]$ enforces the constraint, $<Q>=1$. 
The current is given by~\cite{Meir.92}
\begin{eqnarray}
I(V,T)&=&\frac{2e}{\hbar}\int d \omega \frac{2\Gamma_{L}(\omega)\Gamma_{R}(\omega)}
{\Gamma_{L}(\omega)+\Gamma_{R}(\omega)} \rho(\omega,V,T)\nonumber \\
&\times& [f(\omega+eV/2)-f(\omega-eV/2)].
\label{IVeq}
\end{eqnarray}
%where $\Gamma_\alpha(\omega) = \Gamma(\omega-\mu_\alpha)$ with $\alpha=L,R$. 
The nonlinear conductance $G(V)=dI(V)/dV$ is computed by numerical derivative of the current $I(V)$, whereas the linear-response conductance  
is directly obtained from
\begin{eqnarray}
G(0,T)&=&\frac{2e^{2}}{\hbar}\int d\omega 
\frac{2\Gamma_{L}(\omega)\Gamma_{R}(\omega)}
{\Gamma_{L}(\omega)+\Gamma_{R}(\omega)}
\Big(-\frac{\partial f(\omega)}{\partial\omega}\Big)\nonumber \\
&\times& \rho(\omega,V=0).
\label{GTeq}
\end{eqnarray}
Eqs.(\ref{SFeq3})-(\ref{SFeq4}) together with the Dyson equation for $G(\omega)$ and $D(\omega)$ forms a self-consistent set of integral equations. 
These equations are iterated until a  solution is found with which Eqs.(\ref{DOSeq})-(\ref{GTeq}) can be evaluated.

\begin{figure}[t]
\begin{center}
\includegraphics[width=8cm]{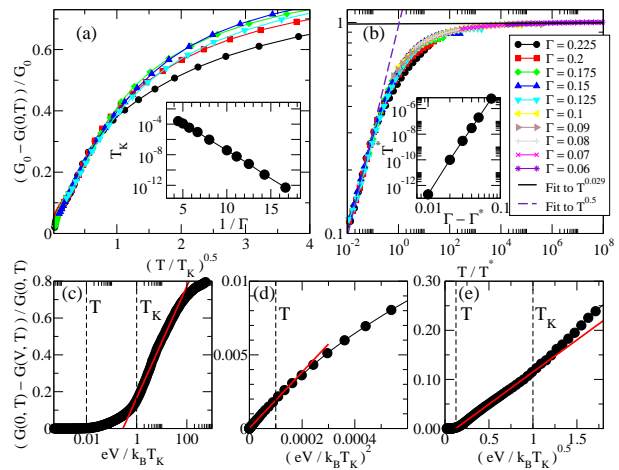}
\end{center}
\par
\vspace{-0.5cm}
\caption{
(a) $(G_0-G(0,T))/G_0$ calculated from Eq.(\ref{GTeq}) displays $T^{1/2}$ behavior for $T<T_K$. 
$G_0\equiv G(0,T_0)_{\Gamma=0.225 D}$ is the linear conductance at 
the lowest numerically accessible temperature $T_0\approx 5\times 10^{-7} D$ at  $\Gamma=0.225 D$. 
Inset: Exponential behavior for $T_K\sim D  
e^{-\pi |\epsilon_d|/\Gamma}$.
(b) An additional power law behavior at high temperature for small $\Gamma\approx \Gamma^{\ast} \approx 0.05 D$, where the crossover scale $T^*$ shows a power law behavior $T^*\sim|\Gamma-\Gamma^{\ast}|^{1/\nu}$ with 
$\nu\sim 0.1$ (Inset). 
Here, $\epsilon_d = -0.2 D$, $\mu =-0.1 D$. 
$(c)\sim (e)$: Non-linear conductance at a large 
$\Gamma=0.2 D$, and fixed parameters: $T_0=5\times 10^{-7}D$, $\mu=-0.1D$, 
$\epsilon_d=-0.2D$. 
The $G(V,T)$ curve shows 2CK Kondo behavior: (c) $log V$ dependence around $V\sim T_K$ with $T_K \approx 5 \times 10^{-5} D$ and $T=T_0$. 
(d) $V^2$ behavior for $V<T$. 
(e) $T^{1/2}$ 2CK behavior for $T<V<T_K$. Here, the red lines are fits to the 
corresponding power-law or logarithmic behaviors in different bias regime.
}
\label{GV_Gamma_exp}
\end{figure}

{\it Results.} 
We now turn to a discussion of the self-consistent solution of
Eqs.~(\ref{SFeq1})-(\ref{SFeq4}); the 
results are summarized in the phase diagram Fig.~\ref{phase}(a).
For simplicity, we focus on the particularly  simple case with parity 
(left-right) symmetry, $U_L=U_R$, $\Gamma_L=\Gamma_R\equiv \frac{\Gamma}{2}$. 

For larger values of 
$j=\Gamma, |\mu|$ (or $j-j^{\ast}>0.1D$)   
with $\Gamma^{\ast}\sim 0.05 D$ and 
$\mu^{\ast}\sim -0.05 D$ 
being crossover scales at a fixed 
temperature $T_0\sim 10^{-7}D$, our results show clear 
2CK behaviors at low temperatures 
$T<T_K$ where $T_K$ is the Kondo temperature defined  as the temperature 
where $G(0,T)$ deviates from a $\sim \sqrt{T}$ behavior\cite{kroha}, 
in agreement with its conventional definition: $G(0,T_K) = G(0,0)/2$.

However, for smaller values of $j$ with $j-j^{\ast}<0.1D$, 
we find universal power-law scaling of $G(T,0)$
distinct from both 
2CK ({\it i.e.} $\sim \sqrt{T}$) and one-channel Kondo ({\it i.e.} $\sim T^2$)~\cite{1CKscaling} 
behavior  at temperatures 
$T\gg T^{\ast}$ where $T^{\ast}$ is the crossover energy scale, 
describing the 2CK-LM crossover. Note that the NCA gives reliable results for $I(V,T)$ even in the single-channel Kondo
case as long as $T \gtrsim 0.1 T_K$.
 
The crossover scale is finite for any non-vanishing gate-voltage ($\mu\neq 0$).
We checked that the crossover scales 
$\Gamma^{\ast},|\mu^{\ast}|$ vanish as $T\rightarrow 0$ in a power-law fashion, 
consistent with the general expectation that Kondo screening in graphene 
can be induced by arbitrarily small doping ($\mu\neq 0$).

The local density of states $\rho(\omega,V)$, given by Eq.(\ref{DOSeq}),  is shown in Fig.\ref{DOSfig}. The Kondo peaks occur
at the chemical potential $\mu$, and the Kondo temperature $T_K$ follows the 
pseudogap Kondo behavior with $r=1$: $T_{K}\sim |\mu| 
\times e^{(-a/|\mu|)}$ (Inset in Fig.\ref{DOSfig}) where 
$a$ has the unit of energy and is a function of the 
bandwidth $D$ and the Kondo coupling $J$\cite{Fradkin}. 
The $|\omega|$ decay in the vicinity of $\omega=0$ is a reflection of the Dirac point of the conduction electrons with DOS $\rho_{c}(\omega)\sim|\omega|$. 
Comparable results have also been discussed in Refs.~\cite{Cornaglia,Vojta2,Xie,Jamal,HGLuo}.

Figs.~\ref{GV_Gamma_exp}(a) and  \ref{GV_Gamma_exp}(b) shows the linear response conductance $G(0,T)$ obtained via numerical derivative 
of Eqs.~(\ref{IVeq}) and (\ref{GTeq}). As long as $\mu \neq 0$, we find clear 2CK behavior at low temperature regime (Fig.~\ref{GV_Gamma_exp}(a)), with $G(0,T)$ displaying 
a $T^{1/2}$ behavior for each coupling $\Gamma$~\cite{kroha}. 
The Kondo temperature $T_K$, 
behaves as: $T_K\sim D  
e^{-\pi |\epsilon_d| / \Gamma}$ (Fig.~\ref{GV_Gamma_exp}(a) Inset). 
For higher temperatures (Fig.~\ref{GV_Gamma_exp}(b)), we find
a universal power-law behavior in $T/T^{\ast}$ for $T>T^{*}$ near 
 $\Gamma^{\ast} \approx 0.05 D$: $G(0,T)\propto (T/T^{\ast})^{\alpha}$ 
with $\alpha\approx 0.029$, indicating the universal crossover 
from 2CK to LM regime. The crossover energy scale $T^*$ shows a 
power-law dependence, $T^*\sim|\Gamma-\Gamma^{\ast}|^{1/\nu}$ 
(Fig.~\ref{GV_Gamma_exp}(b) Inset), where $\nu\sim 0.1$.
\begin{figure}[t]
\begin{center}
\includegraphics[width=0.8\linewidth]{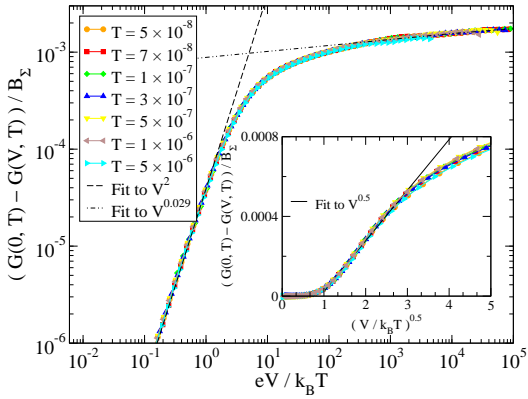}
\end{center}
\par
\vspace{-0.5cm}
\caption{Temperature scaling of the 
nonlinear conductance $G(V,T)$ in the universal crossover region with 
$\Gamma=0.07 D$ and $\mu=-0.08D$ ($\epsilon_d=-0.2D$). 
All curves collapse onto a universal scaling function. It displays $V^{1/2}$ behavior at $T<V<10T$ (Inset). For large $V$ (in units of $D$), $V/T>10^{2}$, $G(V,T)$ shows a power law behavior similar to Fig.\ref{GV_Gamma_exp}(d).}
\label{GV_Gamma_T}
\end{figure}

$G(V,T)$ at larger $\Gamma$ 
shows clear 2CK behavior (see Fig.~\ref{GV_Gamma_exp}(c)-(e) for $\Gamma=0.2 D$). Fig.~\ref{GV_Gamma_exp}(c) indicates the $log V$ behavior predicted for Kondo scattering processes. The 
$V^2$ behavior at $V<T$ and $V^{1/2}$ behavior at $T<V<T_K$ in Fig.~\ref{GV_Gamma_exp}(d) and (e), respectively, are the characteristics of  2CK physics.

\begin{figure}[t]
\begin{center}
\includegraphics[width=0.8\linewidth]{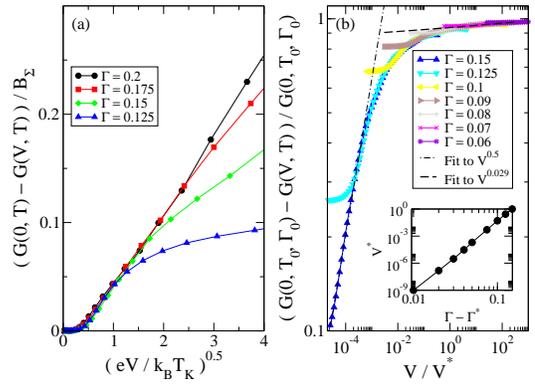}
\end{center}
\par
\vspace{-0.5cm}
\caption{
(a) Scaling of $G(V,T)$ for large $\Gamma$ (parameters  are the same as in Fig.\ref{DOSfig}). 
For $V<T_K$, $G(V,T)$ collapses onto a single curve $\sim T^{1/2}$. 
(b) Power law behavior of  $G(V,T)$
for $V>V^*$. Here, $G(0,T_0,\Gamma_0)$ refers to the linear conductance 
at a fixed temperature $T_0=5\times 10^{-7}D$ and a fixed coupling 
$\Gamma_0=0.2 D$. Inset: Power law behavior of the crossover energy scale $V^*\sim|\Gamma-\Gamma^{\ast}|^{\alpha}$ (see text). 
Here, $\epsilon_d =-0.2D$, $\mu=-0.1D$.
}
\label{GV_GT_mu_scaled}
\end{figure}

To illustrate the crossover between the 2CK and LM regimes, 
$G(V,T)$ is shown in Fig.~\ref{GV_Gamma_T} and 
Fig.~\ref{GV_GT_mu_scaled}.  Fig.~\ref{GV_Gamma_T}, demonstrates $V/T$ scaling~\cite{Kirchner} of $G(V,T)$ for  
$\Gamma=0.07 D$ and $\mu=-0.08 D$. The resulting curves
 collapse onto a universal function. For $V/T>10^{2}$, $G(V,T)$ shows a power law behavior similar to the one shown in Fig.~\ref{GV_Gamma_exp}(b); for 
$1<V/T<10$, the behavior follows  the 2CK $V^{1/2}$ behavior as shown in the inset of Fig.~\ref{GV_Gamma_T}; for $V\ll T$, it shows $V^2$ Fermi-liquid behavior. 
Fig.~\ref{GV_GT_mu_scaled}(a) is the 2CK scaling plot for large $\Gamma$, where  all curves follow the scaling function~\cite{kroha}, $G(0,T)-G(V,T)=B_{\Sigma}T^{1/2}H(A\frac{eV}{T_K})$. Here, $H(x)\propto x^{1/2}$ is a universal function and $B_{\Sigma}$ and $A$ are non-universal constants~\cite{kroha}. 
For $V>V^{\ast}$, 
$G(V,T)$ shows the 
same universal 2CK-LM power-law crossover behavior for small coupling 
$\Gamma\approx \Gamma^{\ast}$ as shown in the linear conductance, $G(0,T)$: 
$G(0,T)-G(V,T)\propto (V/V^{\ast})^{\alpha}$ with 
$\alpha\approx 0.029$,  
(see Fig.~\ref{GV_GT_mu_scaled}(b)). The crossover energy scale $V^*$ also displays a power law behavior $V^*\sim|\Gamma-\Gamma^{\ast}|^{1/\nu}$, where $\nu\sim 0.1$, in line with the equilibrium behavior. 

{\it Discussions and Conclusions.} 
To make contact with a recent STM experiment~\cite{Mattos}, it is necessary to 
generalize our setup  to the case
where one of the leads is made up of a simple ({\i.e.} one-channel) metal  
with constant DOS $\rho_{tip}$ 
near its Fermi energy. 
The ground state in this case will be that of a Fermi liquid. If, however, the tip is coupled only very weakly
($\Gamma_{\mbox{\tiny tip}}/D\ll 1$) to the magnetic adatom on graphene,  the corresponding energy scale will be vanishingly small. In this case it is permissible to replace the normal metal lead by a two-channel lead (with constant DOS at the Fermi energy), as   the RG equations for the one-channel and two-channel Kondo problem coincide in lowest order.
The behavior of $G(V,T)$ in this case are compatible with the results discussed above, see
Figs.~1 and 2 of  Ref.~\cite{note}. 
Our results, see {\it e.g.} Fig.~\ref{GV_Gamma_exp}(c)-(e),  are thus in line 
 with the experimental findings reported in Ref.~\cite{Mattos}.
Fano-lineshapes, derived from our results are shown in~\cite{note}.

Universal out-of-equilibrium scaling is currently pursed in a wide range of condensed matter systems.  As demonstrated, magnetic adatoms on graphene offer the possibility to study steady state properties in a universal crossover regime. 
Here, we have addressed 
the universal crossover regime that separates the local moment regime from the two-channel Kondo regime of adatoms in graphene. Our analysis is based on
the two-channel Anderson model in graphene.  
In particular, we calculated the differential conductance both in the linear and nonlinear 
regime. For sufficiently large hybridization, we found clear two-channel Kondo 
signatures. As the hybridization is reduced, 
the crossover region separating two-channel Kondo and local moment 
ground states is entered and
the crossover  is monitored by the narrowing of
the Kondo resonances.
In the crossover regime, the conductance shows universal power-law behavior. 
Our results not only agree well a recent scanning tunneling spectroscopy experiment but also
provide a comprehensive theoretical analysis  of the transport properties of 2CK impurities in graphene. 

\acknowledgments
We thank T. \ Chowdhury, K.\ Ingersent, J.\ Kroha, K.\ Sengupta, P.\ Ribeiro, F.\ Zamani, M. \ Vojta and in particular H.\ Manoharan for useful discussions. C.H. C. acknowledges the NSC grant No.98-2112-M-009-010-MY3, No.101-2628-M-009-001-MY3, the NCTS, the MOE-ATU program of Taiwan R.O.C.


\begin{thebibliography}{36}

\bibitem{kondo}
A.C. Hewson, The Kondo Problem to Heavy Fermions (Cambridge University 
Press, Cambridge, UK, 1997).


\bibitem{2CK}
P. Nozieres, A. Blandin, J. Phys. (Paris) {\bf 41}, 193 (1980).

\bibitem{Bethe}
N. Andrei and C. Destri, Phys. Rev. Lett. {\bf 52}, 364 (1984), 
 P. B. Wiegmann and A. M. Tsvelik, Z. Phys. B {\bf 54}, 201 (1985).

\bibitem{CFT}
I. Affleck and A. W. W. Ludwig, Nucl. Phys. B {\bf 352}, 849
(1991); ibid. {\bf 360}, 641 (1991). A.W.W. Ludwig, I. Affleck, 
Phys. Rev. Lett. {\bf 67}, 3160 (1991), E. Sela, A.K. Mitchell, L. Fritz, Phys. Rev. Lett. {\bf 106}, 147202 (2011).

\bibitem{bosonization}
V.J. Emery, S. Kivelson, Phys. Rev. B {\bf 46}, 10812 (1992); M. Fabrizio, A.O. Gogolin, Phys. Rev. B {\bf 51}, 17827 (1995).

\bibitem{2CK-NRG}
A. K. Mitchell, E. Sela, D. E. Logan, Phys. Rev. Lett. {\bf 108}, 086405 (2012).

\bibitem{Goldhaber}
R.M. Potok, I.G. Rau, H. Shtrikman, Y. Oreg, and Goldhaber-Gordon, 
Nature London {\bf 446}, 167 (2007).

\bibitem{Cox_a}
D. L. Cox and A. Zawadowski,Adv. Phys. {\bf 47}, 599 (1998). 

\bibitem{Cichorek}
T. Chichorek et al., Phys. Rev. Lett. {\bf 94}, 236603 (2005).

\bibitem{sengupta}
K. Sengupta and G. Baskaran, Phys. Rev. B {\bf 77}, 045417 (2008).

\bibitem{Vojta2}
Matthias Vojta, Lars Fritz and Ralf Bulla, Europ. Phys. Lett. {\bf 90}, 27006 (2010).

\bibitem{Mattos}
L. S. Mattos {\it et al.}, (un-published); 
L.S. Mattos, ``Correlated electrons probed by scanning tunneling 
microscopy'', PhD. thesis, Stanford University (2009).

\bibitem{Fradkin}
David Withoff and Eduardo Fradkin, Phys. Rev. Lett. {\bf 64}, 1835 (1990).

\bibitem{Gonzalez-Buxton.98}
C. Gonzalez-Buxton and K. Ingersent, Phys. Rev. B {\bf 57}, 14254 (1998).

\bibitem{Ingersent}
K. Ingersent and Q. Si, Phys. Rev. Lett. {\bf 89}, 076403 (2002).

\bibitem{Glossop_a}
Matthew Glossop et al., Phys. Rev. Lett. {\bf 107}, 076404 (2011).


\bibitem{Vojta4}
Lars Fritz and Matthias Vojta, Phys. Rev. B {\bf 70}, 214427 (2004).

\bibitem{Zamani.12}
F.\ Zamani, T. \ Chowdhury, P. \ Ribeiro, K.\ Ingersent, and S.\ Kirchner, pss in print (2013) and  arXiv:1211.4450.

\bibitem{Zamani.13}
F.\ Zamani et al., in preparation (2013).


\bibitem{Vojta3}
I. Schneider, L. Fritz, F. B. Anders, A. Benlagra, M. Vojta, 
Phys. Rev. B {\bf 84}, 125139 (2011).

\bibitem{DAnna}
Luca Dell'Anna, J.~Stat.~Mech. P01007 (2010).

\bibitem{GZhu}
Zhen-Gang Zhu, Kai-He Ding, and Jamal Berakdar, Eur. Phys. Lett. {\bf 90}, 67001 (2010).
 

\bibitem{Wehling}
T.~O.~Wehling {\it et al.}, Phys.~Rev.~B.~ {\bf 81} 085413 (2010).

\bibitem{note}
see additional supplemental material.


\bibitem{Cox_b}
D. L. Cox and A. L. Ruckenstein,  Phys. Rev. Lett. {\bf 71}, 1613 (1993).

\bibitem{Meir} 
Ned S. Wingreen and Y. Meir, Phys. Rev. B {\bf 49}, 11040 (1994).

\bibitem{kroha}
Matthias H. Hettler, Johann Kroha, and Selman Hershfield, Phys. 
Rev. Lett. {\bf 73}, 1967 (1994); Ch. Kolf, J. Kroha, M. Ternes, and W.-D. Schneider, Phys. Rev. Lett. {\bf 58}, 5649 (1998).

\bibitem{Meir.92}
Y. Meir and Ned S. Wingreen, Phys. Rev. Lett. {\bf 68}, 2512 (1992).

\bibitem{1CKscaling}
M. Grobis, I. G. Rau, R. M. Potok, H. Shtrikman, and D. Goldhaber-Gordon, 
Phys. Rev. Lett. {\bf 100}, 246601 (2008).

%\bibitem{Chen}
%J-H. Chen et al., Nature Phys. {\bf 7}, 535 (2011).



%\bibitem{Roura}
%P. Roura-Bas, Phys. Rev. B {\bf 81}, 155327 (2010).

%\bibitem{Tan}
%C.L. Tan {\it et al.}, arXiv:0910.5777 (2009).


\bibitem{Cornaglia}
P. S. Cornaglia, Gonzalo Usaj, C. A. Balseiro, Phys. Rev. Lett. {\bf 102}, 046801 (2009).

\bibitem{Xie}
Huai-Bin Zhuang {\it et al.} Eur. Phys. Lett. {\bf 86}, 58004

\bibitem{Jamal}
Ahen Gang Zhu and Jamal Berakdar, Phys. Rev. B {\bf 84}, 165105 (2011).

\bibitem{HGLuo}
Lin Li {\it et al.}, arXiv:1204.2696v1 (2012).


\bibitem{Kirchner}
S. Kirchner and Q. Si, Phys. Rev. Lett. {\bf 103}, 206401 (2009).


\end{thebibliography}
\end{document}

\title{{\bf \Large Supplemental Material}\\
Gate-tuned two-channel Kondo screening by graphene leads:
Universal scaling of the nonlinear conductance}
\author{Tsung-Han Lee$^{1}$, Kenneth Yi-Jieh Zhang$^{1}$, Chung-Hou Chung$^{1,2}$, Stefan Kirchner$^{3,4}$}
\affiliation{
$^{1}$Department of Electrophysics, National Chiao-Tung University,
HsinChu, Taiwan, 300, R.O.C. \\
$^{2}$National Center for Theoretical Sciences, HsinChu, Taiwan, 300, RO.C. \\
$^{3}$Max-Planck-Institut f\"ur Physik komplexer Systeme, 01187 Dresden, Germany\\
$^{4}$Max-Planck-Institut f\"ur chemische Physik fester Stoffe, 01187 Dresden, Germany
}
\maketitle

\begin{figure}[t]
\begin{center}
\includegraphics[angle=-90,width=0.6 \linewidth]{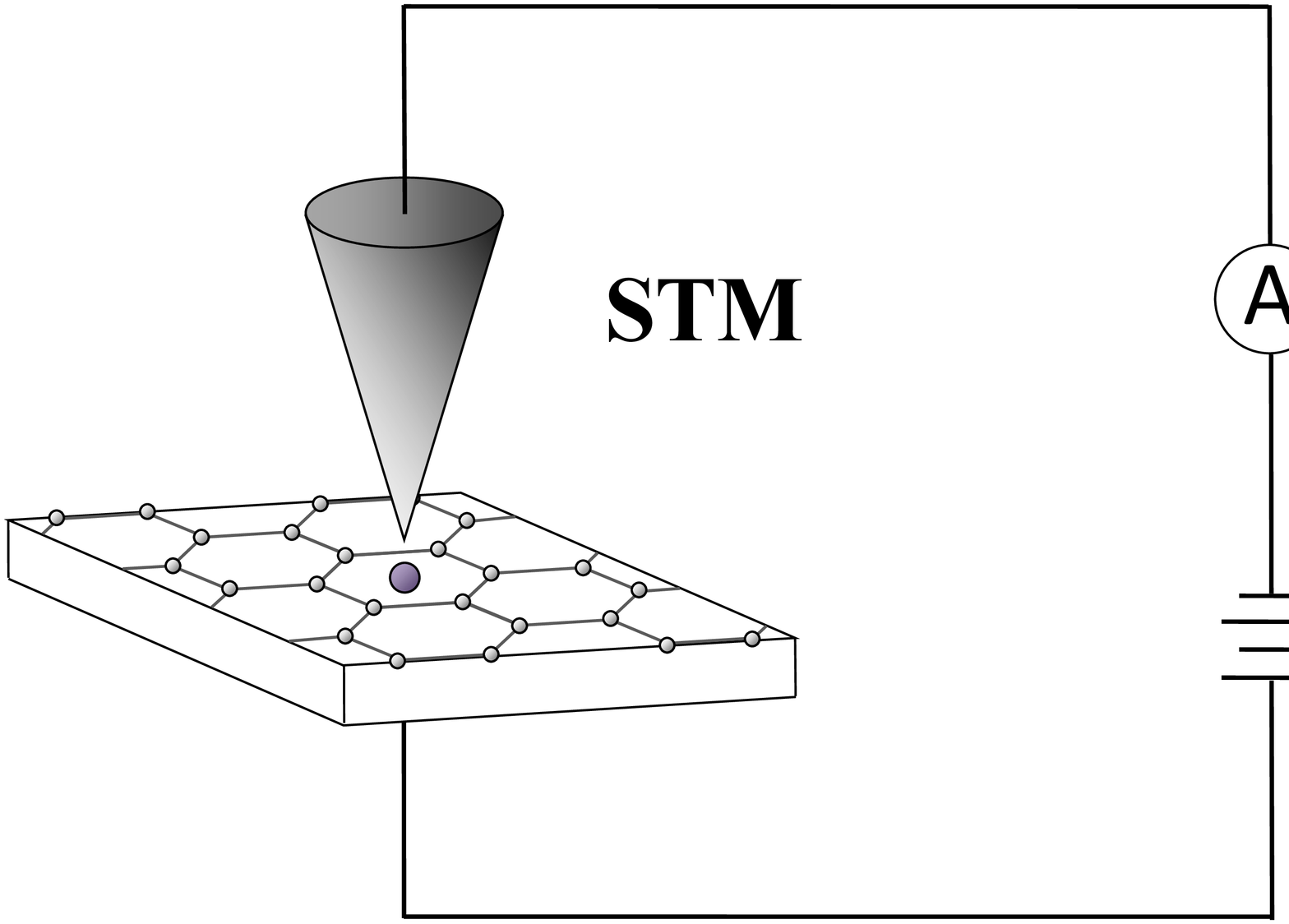}
\end{center}
\par
\vspace{-0.5cm}
\caption{The STM measurement of the magnetic adatom in graphene. 
The $S=1/2$ magnetic adatom is located at the center of the honeycomb lattice 
of graphene.  
}
%\vspace{-2cm}
\label{STM-setup}
\end{figure} 
\section{\bf A: Non-equilibrium NCA}
The extension of the non-crossing approximation onto the Keldysh contour has been discussed 
in several papers~\cite{kroha,Meir}. It is customary to neglect the bias voltage dependence of the conduction electron density of states (DOS) $\rho(\epsilon)$,
This is justified provided $\rho$ is well approximated by a constant in a region around the 
Fermi energy that is large compared to the applied bias voltage. When the DOS vanishes in a power-law fashion at or near the Fermi energy,
this is no longer possible and the equations have to be generalized appropriately.
The full set of equations to be solved for the two-channel pseudogap problem becomes
\begin{eqnarray*}
&& \Sigma_B^<(\omega)=(-2i)\int_{-\infty}^{\infty} \!\!\!\!\!\!d\epsilon \,G_f^<(\epsilon+\omega) \Big [
|V_L|^2 f(-\epsilon+\mu_L) \big . \times  \\
&& \big . \rho_L(\epsilon-\mu_L-\mu)+|V_R|^2 f(-\epsilon+\mu_R)\rho_R(\epsilon-\mu_R-\mu)\Big ]\nonumber \\
&& \Sigma_B^>(\omega)=2i\int_{-\infty}^{\infty} \!\!\!\!\!\!d\epsilon \,G_f^>(\epsilon+\omega) \Big [ 
|V_L|^2 f(\epsilon-\mu_L) \big .\times  \\
&& \big .\rho_L(\epsilon-\mu_L-\mu)+|V_R|^2 f(\epsilon-\mu_R)\rho_R(\epsilon-\mu_R-\mu)\Big ]\nonumber
\end{eqnarray*}
for the pseudo-boson and
\begin{eqnarray*}
&& \Sigma_f^<(\omega)=2i\int_{-\infty}^{\infty} \!\!\!\!\!\!d\epsilon \,G_B^<(\epsilon+\omega) \Big [ 
|V_L|^2 f(-\epsilon+\mu_L) \times \big .  \\
&& \big . \rho_L(-\epsilon+\mu_L+\mu)+|V_R|^2 f(-\epsilon+\mu_R)\rho_R(-\epsilon+\mu_R+\mu)\Big ]\nonumber \\
&& \Sigma_f^>(\omega)=2i\int_{-\infty}^{\infty} \!\!\!\!\!\!d\epsilon \,G_B^>(\epsilon+\omega) \Big [ 
|V_L|^2 f(\epsilon-\mu_L) \times \big .  \\
&& \big . \rho_L(-\epsilon+\mu_L+\mu)+|V_R|^2 f(\epsilon-\mu_R)\rho_R(-\epsilon+\mu_R+\mu)\Big ]\nonumber
\end{eqnarray*}
for the pseudo-fermion. The DOS ($\rho_L$ and $\rho_R$) of  the two leads do not have to be identical. 
The bias voltage applied across the system is 
$eV^{\mbox{\tiny bias}}=\mu_L-\mu_R$, where  $\mu_L$ and $\mu_R$ are the chemical potentials of the left and right leads.

\section{\bf B: Fano-lineshapes}
An experiment reminiscent of the situation considered by us has been performed recently, where 
magnetic adatoms on graphene where investigated via scanning tunneling microscopy (STM), see Ref.~\onlinecite{Mattos}.
Our analysis can be extended to include the current-voltage characteristics measured by an STM (see Fig.\ref{STM-setup}). In this case,
one of the two fermionic leads represents the STM tip and it is necessary to explicitly allow for the different tunneling paths between the STM tip, the
adatom and the substrate which will act as the second lead.  An important difference between the STM setup and our analysis so far is that
the STM tip is a {\it good} metal, {\it e.g.} a single-channel lead with constant DOS at its Fermi energy. We here will model it by a two-channel lead with constant
DOS at its Fermi energy. This is justified provided the coupling between the STM tip and the system is small as the RG scaling equations for two and one-channel
case are identical up to fourth order in the tunneling matrix element.\\
The theory of  STM on 
magnetic adatoms on a metal surfaces has been worked out by Schiller and Hershfield \cite{Schiller} and by
O. \'{U}js\'{a}ghy et al. \cite{Ujsaghy}.
%The characteristic feature of the STM spectra are their Fano-lineshapes \cite{Fano}. 
The current  is obtained from
\begin{equation}
 I(V)\sim \int_{-\infty}^{\infty}d\epsilon \Big[f(\epsilon-eV^{\mbox{\tiny bias}})-f(\epsilon) \Big] \rho_{\mbox{\small tip}} \rho_{\mbox{\small eff}}(\epsilon),
\end{equation}
where $\rho_{\mbox{\small tip}}$ is the density of states of the STM tip and $\rho_{\mbox{\small eff}}$ is an effective density of states probed by the STM and depends on two tunneling rates $t_f$ and $t_c$ that parameterize the hybridization
strength of the STM tip with the magnetic adatom ($t_f$) and the graphene leads ($t_c$). The effective density of states
$\rho_{\mbox{\small eff}}$ can be recast into
\begin{eqnarray}
 \rho_{\mbox{\small eff}}&=&\frac{1}{\pi} \mbox{Im}\Big[
t_c^2 G_c(\vec{R},\vec{R},\epsilon)+(t_d+t_c V G_c(\vec{R},\vec{R}_{ad},\epsilon))\nonumber \\
&\times& G_{ad}(\epsilon)(t_d +t_c V^* G_c(\vec{R}_{ad},\vec{R},\epsilon))
\Big],
\end{eqnarray}
where $V$ is the hybridization strength  between the graphene electrons and the magnetic adatom, $G_c(\epsilon)$ is the advanced local graphene electron 
Green function at the locus of the 
STM tip $\vec{R}$ and $G_c(\vec{R},\vec{R}_{ad},\epsilon))$ is the advanced graphene electron Green function connecting the locus of 
the tip with the position of the adatom at $\vec{R}_{ad}$, and $t_c$ ($t_d$) is the tunneling matrix element between the STM tip and the substrate 
(magnetic adatom).  
$G_{ad}(\epsilon)$ is the advanced Green function of the magnetic adatom that can be obtained from the pseudo-particle Green functions of section {\bf A}.

In the linear regime, the Fano lineshape is given by the 
differential conductance $dI/dV|_{V\rightarrow 0}$, which turns out to be   
proportional to the effective density of states $\rho_{eff}(\epsilon)$:
$dI/dV|_{V\rightarrow 0}\propto \rho_{eff}(\epsilon=V)$.\\ 
$dI/dV|_{V\rightarrow 0}$ can be cast into the Fano lineshape where the Fano parameter q is given by~\cite{Ujsaghy,Fano,chung}
\begin{equation}
q = - \frac{ReG_c^0(\epsilon-{\it i}\eta)}{ImG_c^0(\epsilon-{\it i}\eta)}, 
\end{equation}
and can be treated as approximately constant in the energy range of interest~\cite{Fano}.

\begin{figure}[t]
\begin{center}
\includegraphics[width=\linewidth]{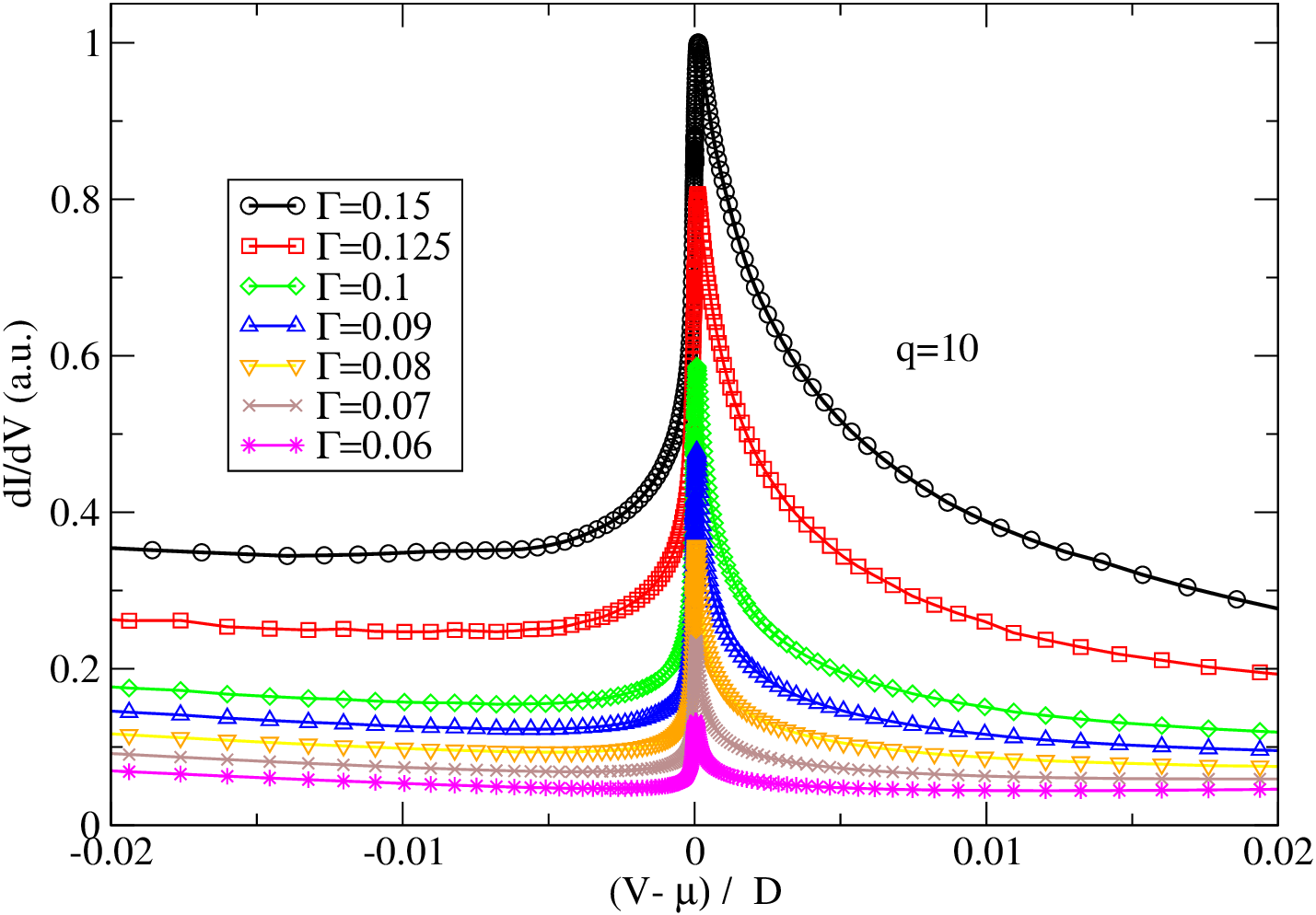}
\end{center}
\par
\vspace{-0.5cm}
\caption{Fano-lineshapes  with Fano parameter $q=10$ for various values of $\Gamma$ (in units of the half-bandwidth
$D$). Here, we have set $\epsilon_d = -0.2D$, $\mu=-0.1D$.}
\vspace{0.5cm}
\label{Fano}
\end{figure}
%opening

\begin{figure}[t]
\begin{center}
\includegraphics[width=\linewidth]{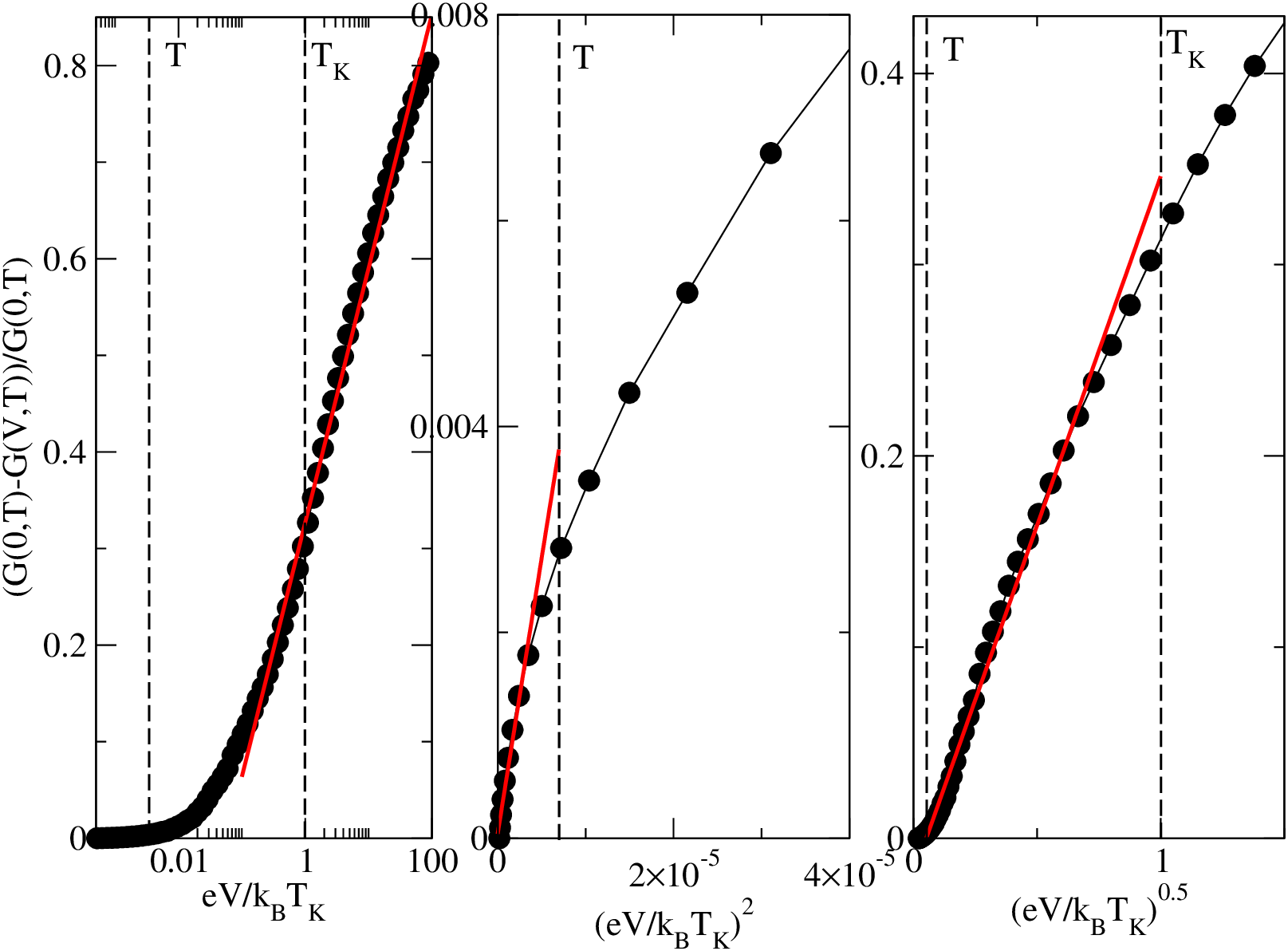}
\end{center}
\par
\vspace{0.0cm}
\caption{Non-linear conductance for a large $\Gamma=0.1 D$ between 
impurity and graphene substrate and a much smaller hoping $\Gamma_{ts}$ 
between tip and the impurity, $\Gamma_{st}\ll \Gamma$,  
corresponding to the STM measurement reported in~\cite{Mattos}. 
The $G(V,T)$ curves agree well with the 
STM results of~\cite{Mattos}. (a) $logV$ dependence around $V\sim T_K$. 
%with $T_K/D \approx 8 \times 10^{-5}$. 
(b) $V^2$ behavior for $V<T$. (c) $T^{1/2}$ 2CK behavior for $T<V<T_K$. Here, 
%$T_K \approx 1.6\times10^{-4} D$, 
$T= 5\times 10^{-7} D$. 
The other parameters are: $\mu=-0.1D$, $\epsilon_d=-0.2D$. }
\vspace{0.5cm}
\label{GV_STM}
\end{figure}

Typical Fano-lineshapes in the linear regime 
are shown in Fig.~\ref{Fano}. 
 The 2CK behavior seen in the STM measurement~\cite{Mattos} 
for Co-adatom at the center of the honeycomb lattice 
is signaled by the Kondo peaks at $\omega=\mu$ in Fano-lineshapes, 
which are compatible with  a large fitting parameter $q$ (for example $q=10$) 
and a correspondingly small $t_c/t_d$ and concomitantly small intervalley scattering.

%The qualitative 2CK behavior of 
%our result for the Fano-lineshapes 
%with $q=10$ is in good agreement with the STM measurement 
%in Ref.\cite{Mattos}. 
%As the hybridization $\Gamma_{L/R}$ 
%is decreased, the Kondo peak gets  narrower and its height at fixed temperature decreases. It finally disappears 
%as $\Gamma\rightarrow \Gamma^{\ast}$.                                              
\begin{figure}[t]
\begin{center}
\includegraphics[width=\linewidth]{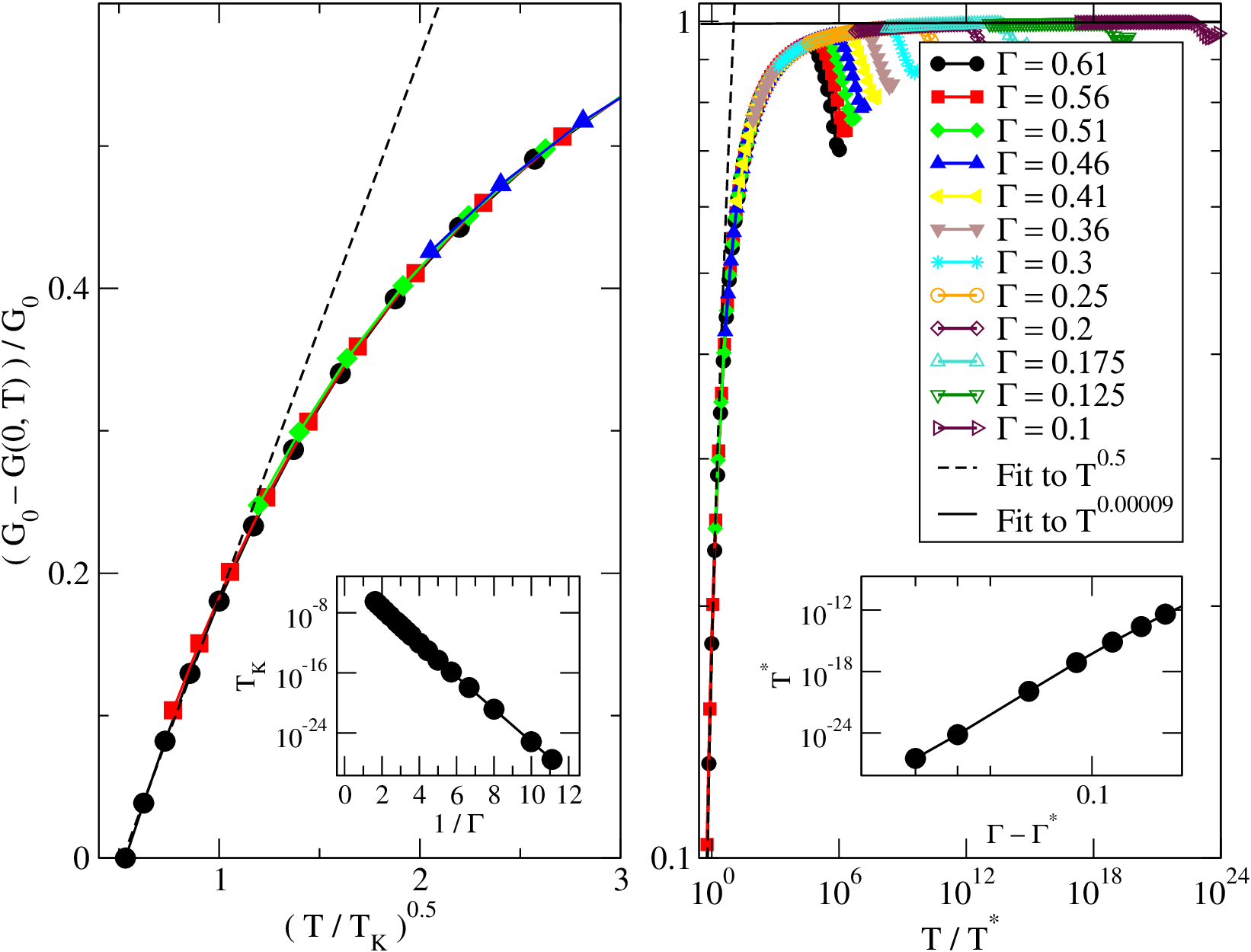}
\end{center}
\par
\vspace{0.0cm}
\caption{Universal scaling in linear conductance $G(T)$ as a function 
of temperature $T$ for at  fixed 
positive chemical potential $\mu=0.1D$ and for various values of 
hybridization $\Gamma$. (a) $T^{1/2}$ 2CK behavior for $T<T_K$ for 
large values of $\Gamma$. 
Inset: $T_K / \Gamma$ vs. 
$1/\Gamma$. 
(b) Universal scaling of $G(T)$ for smaller values of $\Gamma$. 
Inset: The crossover energy scale $T^{\ast}$ as a power-law function of 
$\Gamma-\Gamma^{\ast}$. 
Here, $\epsilon_d=-0.2D$, $G_0$ is the linear conductance for $\Gamma=0.62D$ 
at $T= 5\times 10^{-7} D$, and $\Gamma^{\ast} \approx 0.06D$. }
\vspace{0.5cm}
\label{GT-scaling-pm}
\end{figure}

{\section{\bf C. Universal 2CK-LM crossover for $\mu>0$}

In the main text, we focus on the universal 2CK-LM crossover 
for negative chemical potential, $\mu<0$. 
A similar scaling behaviors can also be found 
in conductance for positive $\mu$. As shown in Fig.~\ref{GT-scaling-pm}, 
for a fixed positive $\mu=0.1D$, the linear conductance $G(T)$ 
vs. hybridization $\Gamma$ follows a 
single universal scaling form of $T/T^{\ast}$ . The single scaling form of $G(T)$ 
we observe here for $\mu>0$ is somewhat surprising as for $\mu<0$ 
the conductance shows two distinct scaling regimes: $T<T_K$ and 
$T>T^{\ast}$. We believe that this difference maybe due to   
the particle-hole asymmetry in our model as the Kondo peak, located 
at $\omega=\mu$, is affected more by the charge peak at $\epsilon_d<0$. 
 for $\mu<0$ than that for $\mu>0$. 

Similar to the case for $\mu<0$, the linear conductance 
for  $\mu>0$ shows a typical 2CK $\sqrt{T}$ behavior for 
$T<T_K$, and a universal power-law behavior at high temperatures 
for $\Gamma\rightarrow \Gamma^{\ast}$: $G(T)\propto (T/T^{\ast})^{\alpha}$ 
with $\alpha \approx 0.00009\approx 0$. 
The Kondo temperature $T_K$ and the crossover scale $T^{\ast}$ 
for  $\mu>0$ behave in a similar way to their $\mu<0$ counterparts:
$T_K \propto \Gamma \times e^{-1/\Gamma}$, 
$T^{\ast} \propto (\Gamma-\Gamma^{\ast})^{1/\nu}$ with 
$\Gamma^{\ast}\approx 0.06D$ and $\nu\sim 0.05$. We believe that our results 
for both positive and negative values of $\mu$ could be used as theoretical 
guidance in future experiments to clarify the issue on two-channel 
Kondo physics in graphene.